# Electrochemical 3D Printing of Silver and Nickel Microstructures with FluidFM

Cathelijn van Nisselroy[a,†], Chunjian Shen[a,b,†], Tomaso Zambelli[a], Dmitry Momotenko[a,c,]*

[a] Laboratory of Biosensors and Bioelectronics, ETH Zürich, Gloriastrasse 35, 8092, Zurich, Switzerland
[b] College of Mechanical and Electrical Engineering, Nanjing University of Aeronautics and Astronautics, No.29, Yudao Street, Qinhuai District, Nanjing 210000, China
[c] Department of Chemistry, Carl von Ossietzky University of Oldenburg, Oldenburg, D-26129, Germany

† Equal author contributions
* Correspondence to e-mail: dmitry.momotenko@uol.de

**Abstract**

Electrochemical 3D printing of conductors with microscale resolution holds a great promise for a wide range of applications, but the choice of suitable metals for these technologies remains limited. Most efforts so far have been focused on deposition of copper, however, other metals are also of interest, especially when tuning of mechanical, electrical, or optical properties is required for a given application. Here we address the issue of a limited materials choice in electrochemical 3D printing by extending the materials library to silver and nickel. Free-standing microscale structures are fabricated in a single step via locally confined electrochemical 3D printing using FluidFM – a microchanneled cantilever nanopipette capable to deliver electrolyte through sub-microscale opening. The 3D printed structures are constructed in a layer-by-layer fashion, which allows complex geometrical shapes such as double rings, helices and tripods. We report the process performance in terms of printing speed (in the range 7 – 40 nm s$^{-1}$ for silver and 27 – 42 nm s$^{-1}$ for nickel) and reveal dense inner structure and chemical purity of the printed features

**Keywords**

## 1. Introduction

Small scale additive manufacturing (AM) relies on a layer-by-layer deposition of material with very high resolution for fabrication of complex objects with micro- and nanoscale dimensions. AM by default permits manufacturing with little to none geometrical constraints, thereby allowing the user to create tailored structures for a wide variety of materials that enable advanced applications like soft-robotic walkers [1], a nervous system on a chip [2], lightweight architectures [3] and hollow glass microstructures [4]. The majority of the 3D printing techniques producing such functional devices process polymer-based materials [5,6]. However, polymers often cannot offer the desired electrical conductivity and mechanical properties required for specific functionality, especially in microelectronics or sensing.

To address these issues, a range of novel metal micro-3D printing strategies has recently been developed [7]. In a short time metal microscale AM has demonstrated itself as a flexible and promising fabrication method in diverse areas, including sensors [8], optics [9], microelectronics [10], microrobotics [11–13], and microelectromechanical systems (MEMS) [14]. Techniques such as direct ink writing (DIW) [14–16], laser-induced forward transfer (LIFT) [17–19], electrohydrodynamic printing (EHDP) [20], and laser-assisted electrophoretic deposition (LAEPD) [21], are transfer-based printing methods, in which the material is pre-synthesized and only needs to be delivered to the specific location. A different approach is taken in synthesis based methods that produce the material on the spot by means of focused electron or ion beams [8,22,23], and electrochemical (EC) printing approaches. The latter family of techniques includes meniscus confined electrodeposition (MCED) [10,24–27], locally confined deposition of precursors in liquid [28–30], localized electrochemical deposition (LECD) [31] and electrohydrodynamic redox printing (EHD-RP) [32].

The range of metals and other materials that are available for above mentioned microscale AM varies widely across different techniques. Transfer-based approaches (DIW, EHDP), relying on extrusion of nanoparticles and/or inks, have only reported the metal deposition of Ag, Au [14–16,20,33], as well as Co and Cu [34], whereas the range of available metals printed by focused electron/ion beams is significantly larger. However, due to the nature of this deposition technique the synthesized metals lack significantly in purity [7], *i.e.* only Pt, Co and Fe were printed with >95% of the main metallic component, whereas other metals (Ti, Cr, Mn, Mo, Sn, Pb, C, Ni, Cu) contain more than 5% of impurities. On the contrary, transfer-based methods offer a broader range of materials: for example, LIFT process has demonstrated the capacity to produce an array of high-purity metals (Au, Ag, Cu, Al, Pt, V, Al, W, Cr, Ni, Cu/Ag alloys [11,17–19,35–38]) and is also suitable for producing structures of complex oxides, superconductors, polymers, biomolecules, and even cells, effectuating printed applications such as biological and chemical sensors [35]. However, this capacity is traded-off by limitations in resolution.

EC micro-3D printing is advantageous since it offers direct single-step fabrication of pure metallic features of arbitrary geometry without the need for support structures and is suitable for printing with nanoscale resolution [27] and impressive complexity [39]. However, most of EC techniques have been employed for 3D printing of copper, which remains the most outspoken metal choice [24,28,30,40]. Copper as it straightforward to plate, as the deposition itself does not evoke unwanted side-reactions and moreover, the range of applicable plating potentials is broad, allowing better control of the printing process. Besides Cu, existing reports describe fabrication of Pt pillars by MCED [41], as well as Cu/Ag multi-metal 3D architectures by EHD-RP [32]. However, EC 3D printing can be extended to a broader range of metals that are in principle suitable for conventional electroplating. As suggested by Hirt *et al.*, this range (potentially available with FluidFM, SICM and MCED approaches), is the largest amongst all of the above discussed techniques [7]. Even though a variety of materials has been successfully electroplated using conventional planar methods, it remains a challenge to adopt these processes to the microscale AM.

The limitation in material choices with regard to EC micro-printing is among most important challenges to be addressed for real-world applications. Though EC 3D printing has proven to yield device-grade copper with notable mechanical properties, which is highly promising for applications requiring high electrical or thermal conductivity [42], it is essential to expand the range of suitable printable materials used in contemporary microfabrication and traditional lithography in order to fulfill its potential. In this work, we explore the capacity of EC micro-printing of silver and nickel.

Silver is known to have a high optical conductivity, related to its inherent dielectric function, thus making it an excellent material for optical and plasmonic applications [43]. For instance, 3D structures consisting of stacked single layers of silver nanoparticles were applied in surface enhanced Raman spectroscopy (SERS) [44], and micron- to millimeter sized silver architectures and radiofrequency devices were printed using DIW [14,33]. In addition, 3D printed silver microstructures are interesting as electrodes for future electrochemical energy storage systems [45]. On the other side, nickel is a common material choice for lithography-electroplating-molding (LIGA) processes as a structural material and molding tool, and is widely used for the microfabrication of MEMS parts *i.a.* micro gears, acceleration sensors, switches and capacitors owing its favorable mechanical and magnetic properties [46,47]. Also, peculiar magnetic characteristics of nickel attract attention due to its potential in future electronics and memory devices: three-dimensional Ni-containing spin-ice nanolattices mimic magnetic monopoles, a rather unusual material, that cannot be obtained in planar geometry [48].

Electrochemical 2D patterning of millimeter-sized nickel structures [49], as well as the additive printing of nanocrystalline nickel thin films with near-bulk material properties [50] have recently been reported. Similarly, by adopting an electroless plating protocol, nickel interconnects and circuit patterns, printed onto commercially available circuit substrates, were created [51]. At the microscale so far only very simple nickel structures (micropillars) have been achieved by localized electrodeposition [52].

The focus of this work is the EC fabrication of 3D nickel and silver microstructures using the FluidFM technology [28,29,53]. Optimization of the printing process in terms of solution composition, deposition potentials and overpressure for localized metal ion delivery allowed control of the feature morphology and print rate. With these optimized printing process conditions, we report fabrication of complex template-free 3D geometries achieved in a layer-by-layer fashion and demonstrate the microstructure and chemical composition of the printed features.

## 2. Material and Methods

*2.1 Chemicals and substrates*

Aqueous solutions were prepared using deionized water with resistivity 18.2 MΩ cm at 25 °C (Milli-Q, Merck). All solutions were filtered through a 0.22 μm PDVF membrane (Millex-VV, Merck Millipore, Germany) prior to printing. For metal printing substrates with 3 nm Ti adhesion layer and 25 nm Au film on glass slides (Paul Scherrer Institute, Switzerland) and prepared in-house on [100] silicon wafers using standard e-beam deposition techniques.

The silver plating ink was based on a commercially available Silveron GT101 low speed plating solution (DuPont, USA) of ~ pH 9.7 with a silver concentration of 15-25 g·L$^{-1}$ (~ 140 – 230 mM). For printing, the ink was diluted 5 times with water (to prevent clogging during Ag deposition), which resulted in a final Ag concentration of ~ 28 – 46 mM. A 150 mM $NH_4OH$ solution (28% $NH_3OH$ in $H_2O$, Sigma Aldrich, USA) of pH ~10.4 was employed as a supporting electrolyte in the bulk. Both the plating ink and supporting electrolyte were filtered through a 0.1 μm and 0.22 μm PVDF membrane (Millex-VV, MerckMillipore, Germany) respectively.

Nickel plating solution contained 600 g·L$^{-1}$ (3.88 M) $NiSO_4$. The supporting electrolyte consisted of 35 g·L$^{-1}$ boric acid (Sigma Aldrich, USA) and 10 g·L$^{-1}$ sodium citrate (Sigma Aldrich, USA) at a pH of 6.

*2.2 Pre-Printing Tests*

Cyclic voltammograms were performed with a scan rate of 0.02 V/s in a range from 0 V to -1.5 V using a PalmSens3 potentiostat with PSTrace 5.8 software (PalmSens B.V., The Netherlands) using Ag/AgCl wire as a quasi-reference electrode and Pt wire as a counter electrode.

*2.3 Preparation of FluidFM probes*

Prior to use standard FluidFM nanopipettes (Cytosurge AG, Switzerland), having a 300-nm aperture at the pyramidal apex and a nominal spring constant of 2 N/m, were plasma treated for 2 min at 18 W (PDC-32G, Harrick Plasma, USA) to facilitate filling of the reservoir with the filtered plating ink.

*2.4 3D Printing Set-Up*

Printing experiments were carried out with two different FluidFM instruments, each having its own EC cell. For deposition of Ni, ink-filled FluidFM nanopipettes were mounted on the head of a FluidFM BOT (Cytosurge AG, Switzerland) which was installed on top of an inverted microscope (Zeiss Axiovert 200, Zeiss, Germany). A cleaned Au substrate (working electrode WE) was placed inside the printing chamber (the 3-electrode EC cell, Cytosurge AG, Switzerland), which consisted of a Teflon plate holder, a graphite counter electrode (CE) and an Ag/AgCl quasi-reference electrode (RE). All electrodes were connected to an external potentiostat (PalmSens3, PalmSens B.V., the Netherlands). To pre-fill the cantilever with plating solution, an overpressure of 500 mbar (supplied by the internal BOT pressure controller) was applied to the probe. In contrast to other microscale electrochemical 3D printing methods, relative humidity has no influence on the printing process since electrolyte evaporation from the cantilever does not occur (as it is not exposed to ambient environment) and the change of concentration of electrolyte in the bath is close to negligible within the course of printing experiments. As soon as the liquid ink was ejected from the cantilever tip, the printing chamber was manually filled with the supporting electrolyte and the probe was lowered into the bath. During the printing process, the instrument followed pre-defined voxel coordinates loaded into the printing software (CAPA, Exaddon AG, Switzerland). For printing Ag, a another FluidFM setup was employed. In this custom-built system FluidFM nanopipettes were mounted onto a Nanowizard 1 AFM (JPK Bruker, Germany) which was installed on top of an inverted microscope (Zeiss Axiovert 40 MAT, Zeiss, Germany). Printing was performed inside a home-made printing chamber, consisting of an Au substrate (WE), a platinum wire CE and an Ag/AgCl wire as quasi-reference. During the printing process the pressure was supplied by an external pressure controller MCFS 4C (Fluigent, France). The instrument followed pre-defined voxel coordinates read from a file by the custom printing program written in LabVIEW (National Instruments, USA). Individual voxel height was set to 200 nm, even though printing at 250 nm or 500 nm would also be possible. Voxel heights >500 nm will cause less confined metal deposition, whereas heights <200 nm are prone to nozzle clogging. The difference in configuration between printing Ag and Ni was dictated exclusively by the availability of the instruments, as there is no fundamental advantage offered by one or another instrument for printing either of the metals.

*2.5 Electron Microscopy Analysis*

The samples were analyzed in a JSM-7100F SEM (JEOL, Japan) using a 10 - 15 kV acceleration voltage and a sample tilt varying between 0° and 60°. EDX spectroscopy (AMETEK-EDAX EDX detector, EDAX TEAM software) was run in point analysis mode on each sample with an acceleration voltage of 15 kV and no sample tilt (perpendicular to the substrate). For microstructural analysis a Helios 5 UX DualBeam FIB-SEM (Thermo Fisher Scientific, USA), was used to expose samples' cross sections as well as image them. Milling of the structures was performed with a 41 pA (Ag) or 0.26 nA (Ni) current for coarse cross sections, followed by polishing step of 26 pA (Ag) or 41 pA (Ni). Secondary electron images of the cross sections were obtained at an acceleration voltage of 5 kV for Ag in immersion mode, whilst the Ni structures were imaged at 2 kV in normal SEM mode. Both Ag and Ni SEM images were obtained with tilt correction.

## 3. Results and Discussion

*3.1 FluidFM 3D Printing Principle*

FluidFM is an atomic force microscope (AFM) adopting a cantilever with a hollow microfluidic channel that is connected to closed fluid reservoir and an external pressure controller [53]. For printing, the microfluidic channel is filled with the desired metal salt electrolyte (ink) and mounted to the AFM head [28,29,53]. The cantilever enables local delivery of metal ions into the specified locations on the conductive substrate (working electrode), placed inside an electrolyte filled electrochemical cell. The choice of the substrate material depends on the application and the material to be electrodeposited, but in general, common electrode materials, such as metals (Cu, Au, Pt), non-metals (carbon), metal oxides (indium tin oxide) and semiconductors are suitable for local electrodeposition. Metal ion reduction occurs on the working electrode area directly under the cantilever's apex. As a result, a solid voxel is growing out-of-plane until it touches the cantilever's apex causing its deflection, which is detected by a reflected laser beam on a photodetector. Importantly, the small cantilever deflection has negligible influence on electrolyte delivery and the overall printing process: feedback allows detection of touch events within ms time scale, while the metal growth is usually 2-3 orders of magnitude slower than that. This AFM feedback enables fully automated printing: each touching event indicates a completion of a volumetric voxel, which is followed by a software command to translate the nanopipette to the next printing location. This allows reaching a full layer-by-layer printing capacity. By changing printing parameters such as the overpressure, plating potential and aperture diameter of the FluidFM probe, the size of each individual voxel can be tailored by an order of magnitude [29]. This interplay between different process parameters can be assessed quantitatively using well-known theory developed for a so-called "wall-jet electrode"[54], since FluidFM is a special case of this electrochemical system [29]. A more extensive description of the printing instrumentation and procedures can be found elsewhere [28,29].

*3.2 Pre-Printing Experiments*

Prior to printing at the microscale, EC reduction of silver and nickel was first macroscopically investigated and the composition of the supporting electrolytes was then adjusted accordingly. For this purpose, cyclic voltammetry (CV) was employed to provide further insights into the electrochemical processes of metal reduction.

Silver electrodeposition CV from the diluted Silveron GT101 solution is shown in Figure 1a. The key for optimizing Ag plating process was the selection of supporting electrolyte. The exact contents of the commercial Silveron GT101 is undisclosed, but the ink pH value (~pH 9.6) suggested the use of 150 mM $NH_4OH$ solution (~pH 10.5) as a suitable supporting electrolyte. Silver forms soluble ammine complexes thus allowing adjustment of metal concentration without possible Ag losses due to AgOH precipitation. Essentially featureless CV (dotted line in Figure 1a) indicates that the chosen supporting electrolyte is electrochemically inert within the selected voltage window. Addition of the Silveron GT 101 (to the final Ag concentration ~ 28 – 46 mM) leads to the appearance of a cathodic process with the onset of silver reduction at ca. -0.25 V. At potentials below -1 V hydrogen evolution reaction starts making a significant contribution to the cathodic current. Based on these observations a plating window within -0.5 V to -1 V was chosen for this silver solution.

For nickel printing a solution of 3.88 M $NiSO_4$ also containing 566 mM boric acid and 39 mM sodium citrate with a pH of ~6 was used. The choice of these additives for Ni plating is dictated by several factors. Normally, Ni plating is performed in aqueous electrolytes at ~pH 4, however, at this pH hydrogen evolution reaction leads to intense bubble formation, which disturbs the laser-based AFM feedback employed in FluidFM. Increasing the pH to higher values diminishes the intensity of hydrogen evolution reaction, but brings another issue, since at pH 6, there is a considerable probability to form nickel hydroxides during electroplating due to local pH changes at the cathode surface. This effect is mitigated by the addition of boric acid which locally buffers the solution. The role of sodium citrate is to form complexes with nickel ions, which adsorb on the electrode and help achieving smoother surface morphology of the deposit. Figure 1b shows the CVs for the supporting electrolyte and the $NiSO_4$ ink solution. Essentially flat voltammogram (dotted line in Figure 1b), representing the supporting electrolyte without metal ions, indicates the absence of electrochemical reactions occurring in the buffer within the displayed voltage window. The reduction of $NiSO_4$, here displayed using a solid line, has an onset potential at ca. – 0.43 V. However, visual changes in the electrode appearance occur only around ca. -0.8 – -0.9 V. Due to the limitations imposed by the hydrogen evolution reaction at voltages < -1.0 V, a similar printing window as for Ag ( -0.5 V to -1.0 V) was chosen.

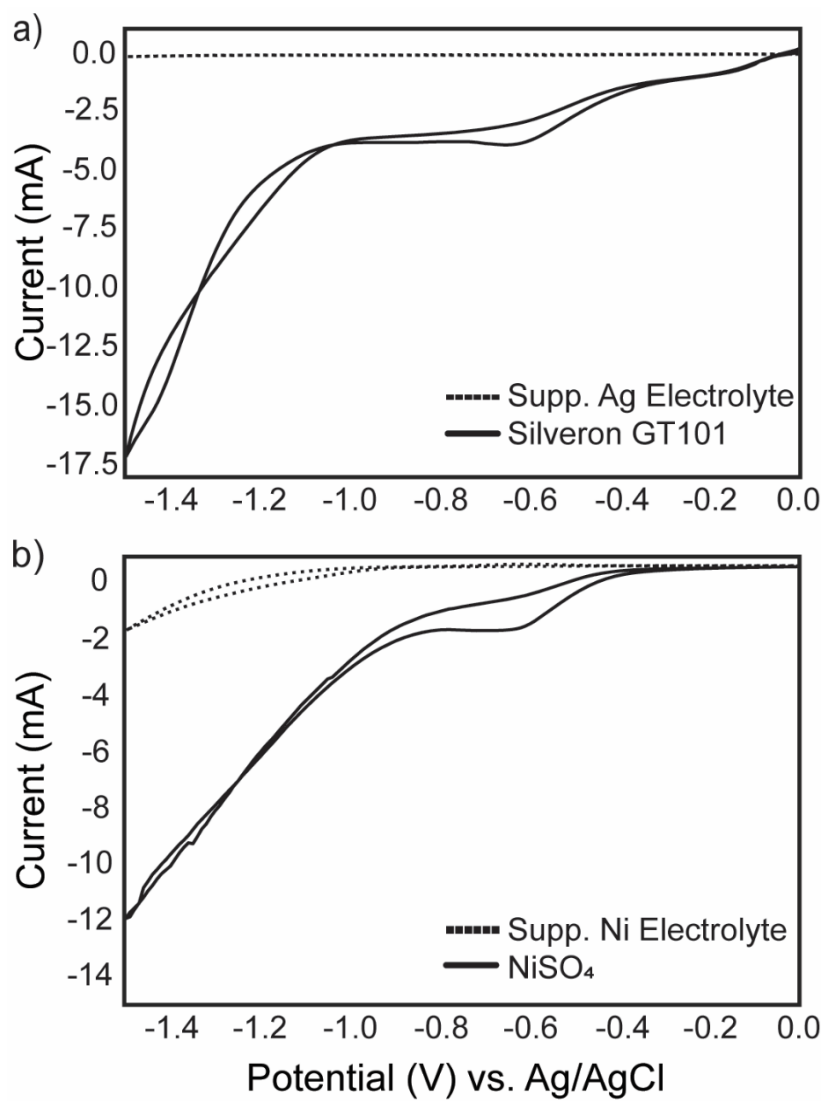

*Figure 1.* Voltammetric study of the plating inks (solid lines) and supporting electrolytes (dotted lines) for **(a)** diluted Silveron GT101 silver plating solution (~ 30 – 50 mM) and **(b)** Ni ink.

*3.3 Optimization of the microprinting process*

Macroscopic electrochemical studies provide the information on the deposition process itself and help choosing suitable electrolyte chemistry and voltage window, but microprinting requires finer tuning of the process parameters. Among the most important is achieving high print rates, while keeping the control of the size and morphology of the deposited features. It is thus crucial to adjust plating voltage (within the chosen window identified in the CV experiments) since it also affects the morphology of the printed metal, and ensure a proper choice of other parameters, such as ink concentration and overpressure for localized delivery. When tuning the concentration of metal in the ink solution, at a first glance, the choice is rather obvious: a higher amount of salt results in faster printing. However, very fast (or uncontrolled, sporadic) growth is strongly undesired, since it can result in a complete or partial nozzle clogging. This seemed especially pronounced in the case of Ag deposition. Only after dilution of the original Silveron GT101 by a factor of 5, resulting in Ag concentration of ~ 30 – 50 mM (as in the preliminary macroscopic voltametric investigations), a stable printing unaffected by nozzle clogging was observed. On the contrary, the nickel ink did not show any issues with nozzle clogging even with high metal content.

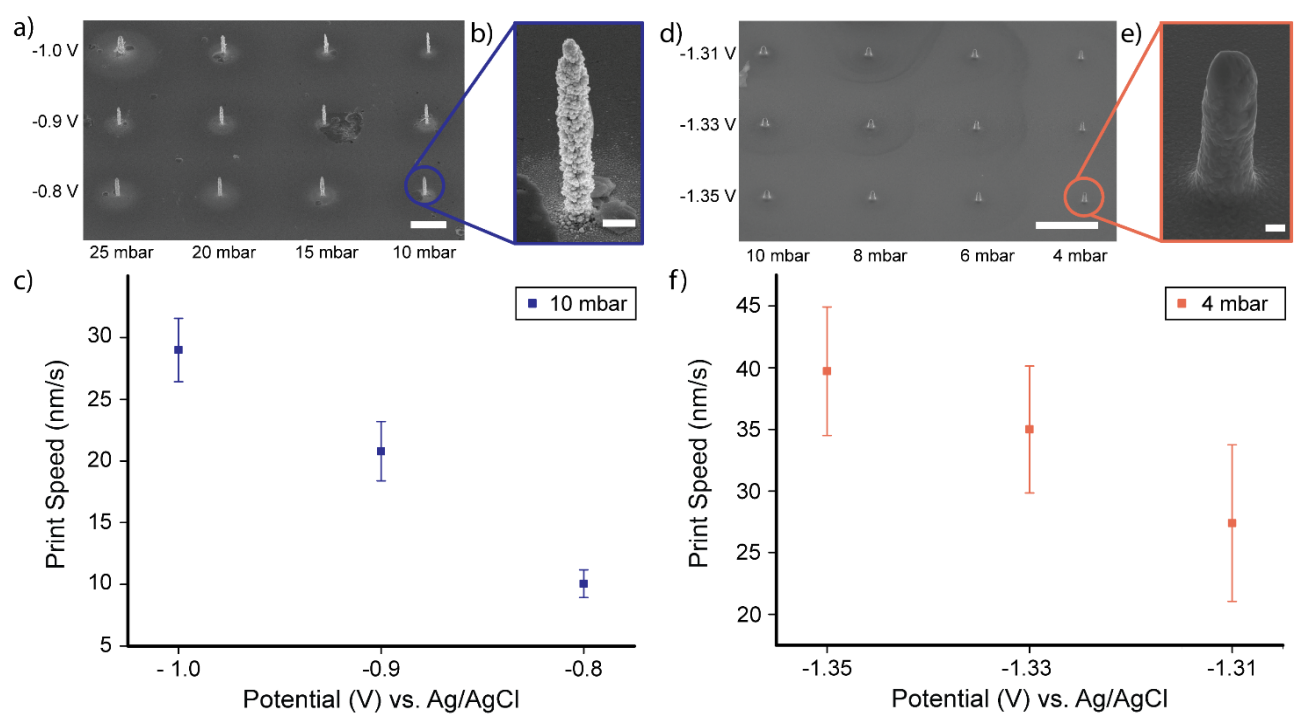

*Figure 2.* Investigation of the process parameters on the EC printing of metals. SEM images of **(a, b)** Ag and **(d, e)** Ni micropillars. **(b, e)** Zoomed SEM micrographs of individual Ag (printed at -0.8 V and 10 mbar) and Ni (printed at -1.35 V and 4 mbar) pillars, respectively. **c, f)** Influence of the applied voltage on the print rate for Ag pillars deposited at 10 mbar **(c)** and Ni at 4 mbar **(f)**. Statistical data is calculated with N = 3. Scalebars 10 and 50 μm for graphs (a) and (d), and 1 μm for (b) and (e). SEM micrographs are acquired at 45° tilt angle.

Optimization of other parameters was achieved by printing micropillar arrays (Figure 2). Each array consisted of 12 pillars: 3 rows with varying deposition potential and 4 columns with varying overpressure. Each pillar was fabricated until the total height of 10 μm and was printed as a stack of 50 voxels, each 200 nm tall (nominal). A smaller voxel height value could lead to unneccesary slower printing (more time spent on translation between voxels). On the other hand, when individual voxels are much taller, the increased distance between the tip and feature results in a less confined ion supply that compromises collection efficiency by the substrate and slows down the process. Here, the printed Ag pillars (Figure 2a,b) exhibit an actual total height of 8.31 ± 0.36 μm with diameters ranging from 1.49 ± 0.16 μm (25 mbar) down to 0.95 ± 0.2 μm (10 mbar). The Ni pillars (Figure 2d,e) reached a height of 8.39 ± 0.45 μm with diameters between 3.77 ± 0.16 μm (10 mbar) and 2.49 ± 0.1 μm (4 mbar).

Variation of the voltage have shown that the print rate at the lower end of cathodic plating potential window is unacceptably slow, dictating the choice of more negative voltages than could be expected from the voltammetric data shown in Figure 1. From such experiments the print rate of each pillar can be calculated by dividing the actual height of each pillar by its total printing time. Figure 2c demonstrates that, for Ag pillars deposited at 25 mbar, at -1.0 V the growth reaches 29.0 ± 2.6 nm s$^{-1}$, which falls to 20.8 ± 2.4 nm s$^{-1}$ at -0.9 V and then drops to 10.1 ± 1.1 nm s$^{-1}$ already at -0.8 V. Lower cathodic potentials make the printing process even slower to some impractical values, where each voxel requires more than 100 seconds to be completed. Similar trends are noticed for Ag pillars at different pressures (Supporting Information Figure S1a). Interestingly, regardless the pressure value at -1.0 V the print rate shows large variation that indicate instability in the printing process, especially when compared to printing results at -0.8 V and -0.9 V. This could be rationalized by the difference in morphology, dictated by the growth mechanism of Ag, which at -1.0 V yields highly branched dendritic structures that may complicate the AFM laser deflection (*vide infra*).

Nickel printing process exhibited a similar trend with voltage, however, growth rate at potentials > -1.31 V becomes very sluggish, while deposition at < -1.35 V results in perturbations arising from hydrogen bubbles. Within this narrow voltage window (-1.35 to -1.31 V) the printing is stable and shows steady growth rates varying from 27.40 ± 6.34 nm s$^{-1}$ at -1.31 V to 39.71 ± 5.20 nm s$^{-1}$ at -1.35 V for pillars printed at 10 mbar (results with other pressures in Supporting Information Figure S1b). As expected, also for Ni increased growth rates are found at more negative printing voltages. In fact, the effect of the potential on the growth rate for Ni is around 2.5 - 3.5 times larger than for Ag. For Ni deposition rate increases with 5 to 7 nm s$^{-1}$ per 20 mV of more cathodic voltage, whereas it is only 10 nm s$^{-1}$ per 100 mV for Ag.

These experiments showed, that the vertical print rates are about one order of magnitude lower as compared to our previously reported results for copper (250-500 nm s$^{-1}$ for a similar pressure range) [28]. In silver printing, this may be related to a 21 – 35 times lower ion concentration, while sluggish nickel plating is likely to be related to a slower intrinsic kinetics of the process. Compared to other elecrochemical techniques, the print rates in this work are slightly slower than those reported for Cu deposition from glass nanopipettes (115 nm s$^{-1}$) [29] and about 25-100 times lower as reported for EHD-RP (1 μm s$^{-1}$) [32].

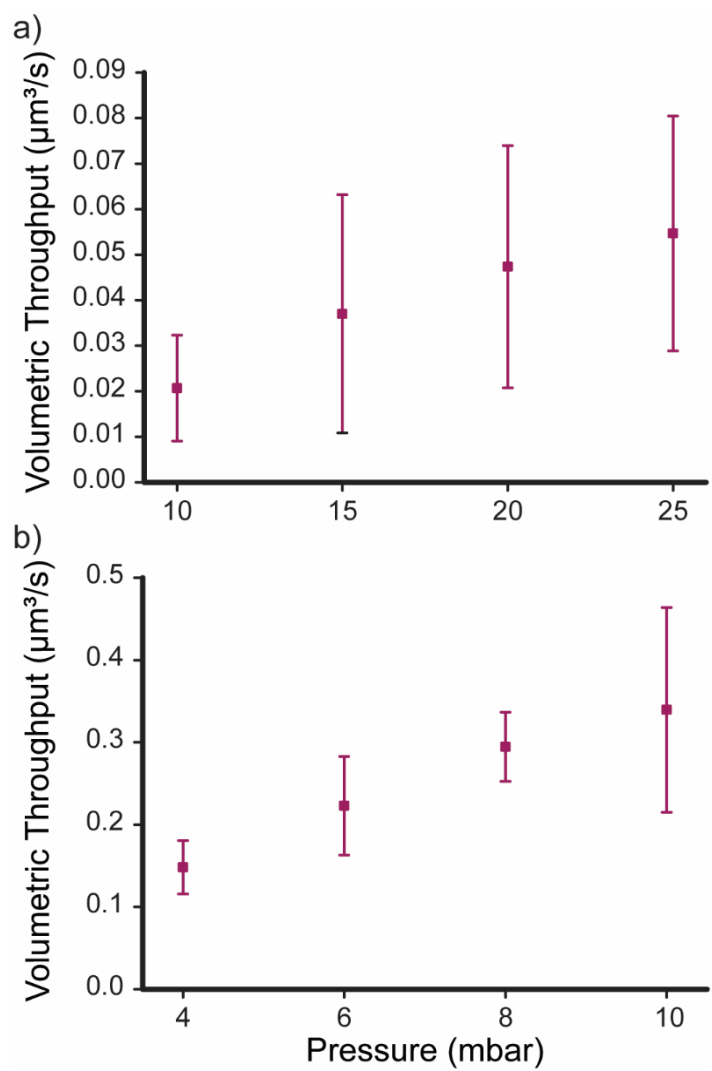

*Figure 3.* Volumetric print rates of the fabricated pillar arrays shown in Figure 2 for **(a)** Ag within overpressure range of 10 to 25 mbar at -1.0 V and **(b)** Ni (4 to 0 mbar) at -1.35 V. Standard deviation and the mean values are calculated for N = 3.

The overpressure for precursor delivery also plays an important role for metal printing. The chosen overpressures for Ag were 10 – 25 mbar (with step of 5 mbar between pillars) and for Ni 4 – 10 mbar (pressure step 2 mbar). Since pressure also significantly affects the pillar diameters, it is convenient to represent feature growth in terms of volumetric throughput in $\mu m^3$ $s^{-1}$ (Figure 3), which is defined as the total volume of a pillar divided by the print time, required for its completion. Figure 3a shows the volumetric print speed for Ag pillars fabricated at -1.0 V, starting with $0.021 \pm 0.012$ $\mu m^3$ $s^{-1}$ at 10 mbar, increasing to $0.037 \pm 0.026$ $\mu m^3$ $s^{-1}$ at 15 mbar, up to $0.047 \pm 0.027$ $\mu m^3$ $s^{-1}$ at 20 mbar, up to a final value of $0.055 \pm 0.026$ $\mu m^3$ $s^{-1}$ at 25 mbar. Printing at -0.9 V and -0.8 V show similar trends (see Supporting Information Figure S2a). The volumetric throughput for Ni, shown in Figure 3b, exhibits analogous tendency with the lowest throughput of 0.181 ± 0.075 µm$^3$ s$^{-1}$ reported for 4 mbar up to 0.417 ± 0.150 µm$^3$ s$^{-1}$ for 10 mbar for pillars printed at -1.35 V (for other printig potentials see Supporting Information Figure S2b). From Figure 3 it can be noticed that for both Ag and Ni the total volumetric throughput increases with increasing pressure. For copper it increases linearly within 0 – 100 mbar and then levels-off at larger pressure values [29]. Here, however, it is difficult to univocally confirm this trend because of the large printing instability at higher overpressures as evident from the uncertainty in print rates. For Ag this instability can arise from approximating the pillars' morphology to a cylinder (while in reality, the shape is more complex given the roughness of the surface). The relative variation in print rates for Ni appears to be lower, but the reason for this is unclear. When comparing the volumetric throughput with those of other micro- and nanoscale printing approaches, the observed values are tens of orders of magnutide lower than reported for LIFT [7], but are comparable to post-processed EHD (for Ag) [55] and focused electron beam deposition (for Ni) [56].

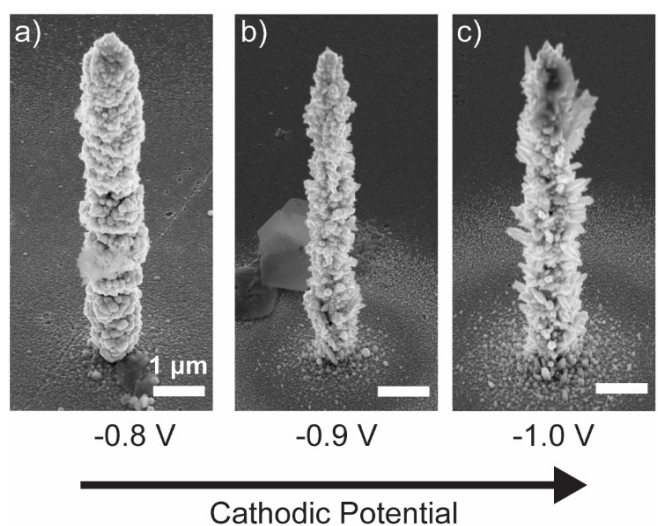

*Figure 4* **a-c)** SEM images of Ag pillars printed in 150 mM NH$_4$OH showing the wide variety of possible morphologies. Pillars are printed at different overpotentials with an overpressure of 10 mbar. Scalebar 1 µm, image tilt: 45°.

Another important aspect is the surface morphology of the features. According to our results, the overpressure has little influence on the roughness of the deposits (at least within the chosen range of values), however, the applied potential has a dramatic effect. Figure 4 shows that for Ag printing more cathodic voltages lead to dendritic feature growth: the relatively smooth pillars at -0.8 V (Figure 4a) change to rough and spiky surface finish with multiple "branches" at -0.9 V (Figure 4b), and this effect further intensifies at -1.0 V (Figure 4c). This branching effect has also been reported previously for copper electrodeposits [57]. In contrast to the Ag sample, the surface of the Ni pillars samples is much smoother. Typically, the Ni electrodeposition is characterized with a low exchange current density (in contrast to Ag) that leads to the formation of smooth deposits across a wider potential window [58].

*3.4 3D microobjects*

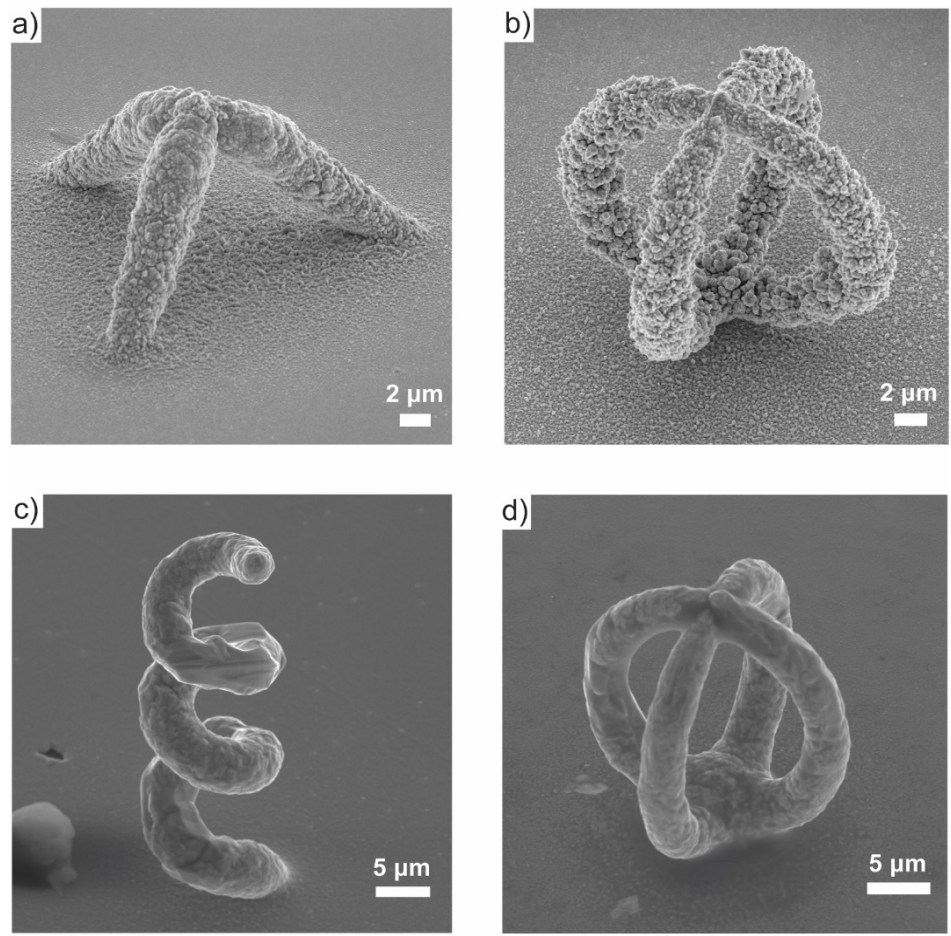

**Figure 5.** SEM images of the 3D printed architectures demonstrating the layer-by-layer nature of the printing process. **a)** Leaning pillars at 45° forming a tripod structure of ~ 17 μm and **b)** double rings of ~ 27 μm, both printed with Ag. Ni single helix **(c)** of ~ 45 μm and a double ring structure **(d)**. Micrographs are acquired at 60° (a) and 45° (b,c,d) tilt angles.

Following the optimization of the microprinting process for Ag and Ni, the creation of complex 3D geometries was demonstrated. Most state-of-the-art micro-3D printing strategies can effortlessly fabricate simple in-plane pillars, but the realization of more complicated 3D objects with out-of-plane details is often challenging. Automated synchronization of the nozzle movement with feature growth appears to be an efficient strategy to print a wide range of geometries, including inclined and overhanging structures in a layer-by-layer fashion. The integrated force-feedback of the FluidFM allows this by detecting the completion of each voxel's printing.

To demonstrate FluidFM's capacity in fabrication of complex features with Ag and Ni, structures such as a tripod, a single helix and a double ring were printed (Figure 5). All these designs contain inclined voxels (that could be also placed within a range of angles with respect to the substrate, like in a double-ring). Importantly, the tripod and double ring structures could be fabricated only by the layer-by-layer printing. A particular geometric inherent difficulty is the top part of the part of the tripod and the double ring designs, that connects all parts of the feature in a single point. This requires precise placement of all voxels that form the structure, achieved by optimizing the lateral voxel spacing, voxel height , and taking into account possible overlap of the neighboring voxels.

Figure 5 shows a ~17 μm-tall tripod Ag structure, formed by three pillars tilted at 45°. With optimized printing parameters, such as overpressure of 15 mbar, voxel height of 200 nm, and a deposition voltage of -0.9 V, the fabrication required ca. 120 minutes to accomplish. The ~ 27 μm-tall double ring geometry (Figure 5b) with a total diameter of 20 μm required slightly shorter time (close to 100 minutes), despite a larger number of voxels forming this structure (683 for double ring vs 319 for tripod). Similar nickel microobjects were fabricated with an overpressure of 6 mbar and a deposition voltage of -1.33 V (helix, Figure 5c) and -1.35 V (double-ring, Figure 5d). The helix with a total height of ~ 44 μm and diameter of 15 μm was printed in 40 min, while the double-ring (same geometry as for Ag) took about 55 minutes. Faster Ni printing is dictated by a higher volumetric throughput of Ni vs Ag (*vide supra*). Interestingly, printing times for both structures are considerably longer than for much simpler pillar arrays because, for the majority of the voxels, the growth direction, as well as the metal salt supply, are out-of-plane, while the pillar voxels grow into the direction of the ion flow. For both metals, one can furthermore notice that the deposition occurs not only directly on the structure but also on the substrate around it. This effect is a well-known characteristic of printing with FluidFM, which occurs inevitably due to ion diffusion inside electrolyte solution.

*3.5 Characterization of the internal morphology and composition*

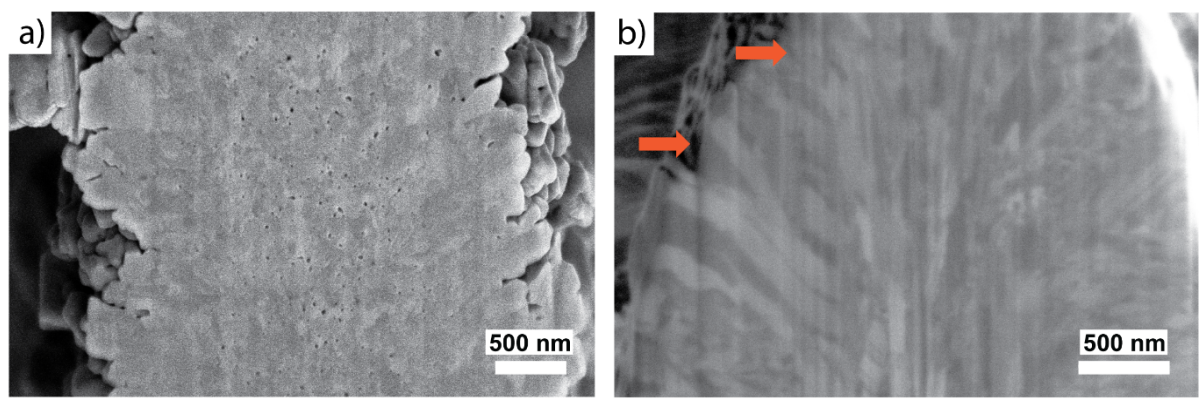

*Figure 6*. Silver **(a)** and nickel **(b)** FIB-SEM cross section images of the double rings structures taken along the printing direction The vertical lines indicated by the orange arrows are an imaging artefact consequent to the FIB-related curtaining effect. Images are obtained at 52° and are tilt corrected.

To investigate the inner structure of the printed metal features, cross-sections of the Ag and Ni double rings (Figure 6a and b, respectively) were exposed by the focused ion beam (FIB). Silver shows a dense, polycrystalline profile with grain sizes in the nanometer range. (ca. 50 – 150 nm). In the central part of the cross section in Figure 6a, however, several nanometer sized voids (<50 nm) are visible. Most likely, they emerged due to trapping the hydrogen nanobubbles formed during printing. Inner structure of Nickel in Figure 6b reveals a comparable microstructural density, but indicates the absence of voids. Yet, a distribution in grain size and shape is noticeable and appears more directional: the grains are elongated and extend from the feature center outwards, resembling a dendritic structure, which, however, does not result in a rough surface of the features. Overall the described internal structure of both printed Ag and Ni shows a dense and crystalline profile as has been reported earlier for EC printing techniques [42]. Nonetheless, the Ag sample has an internal inner structure which rather resembles the Cu printed by EHD-RP, whereas the Ni sample has features more similar to Cu printed by FluidFM.

Additionally, EDX analysis of the chemical composition of the printed features was performed. The spectra (see Supporting Information Figure S3) indicate high purity of both metals. In particular, the printed Ag pillar appears impurity free, as indicated by the absence of major peaks of other elements, with all characteristic Ag peaks clearly identified. On the Ni sample the EDX shows the presence of Ni on both the substrate and the printed pillar, caused by a presence of a thin layer of Ni that was pre-electroplated on the substrate to facilitate further Ni printing.

## 4. Conclusions

In this work we presented the single-step electrochemical 3D printing of microscale silver and nickel structures using FluidFM. Owing to the system's integrated force feedback, the electroplated metal voxels could be deposited individually in a layer-by-layer fashion, yielding elaborate freestanding 3D geometries like the single helix and the double ring. Pillar arrays of the respective metals were printed with various pressures and overpotentials, resulting in vertical printing speeds in the range of 7 - 40 nm s$^{-1}$ (Ag) and 27 – 42 nm s$^{-1}$ (Ni). Dense and polycrystalline microstructures for both metals were observed on the cross sections of the printed structures. Qualitative EDX spectroscopy spectra indicated nearly contaminant-free metals which, along with the prominent microstructural results, confirms the high quality nature of the printed silver and nickel. Although in this work electrochemical printing of only silver and nickel is shown, we envision further advances to optimize other electroplatable metals for microprinting, including noble (gold, platinum, palladium) and non-noble metals (tin, iron), as well as exploring multi-metal printing capabilities in the future, thus extending the technology towards multi-material additive manufacturing at the microscale.

**Acknowledgements**

This work was funded by Innosuisse (Swiss Innovation Agency) under project 18511.1 PFNM_NM (to TZ), the China Scholarship Council No. 201706830034 (C.S.) and has received funding from the European Research Council (ERC) under the European Union's Horizon 2020 research and innovation program (Grant agreement No. 948238, D.M.). We acknowledge Giorgio Ercolano and Wabe Koelmans from Exxadon AG (Switzerland) for their support and advices. The authors thank Stephen Wheeler and Aldo Rossi (ETH Zurich, LBB) for technical assistance. For the support and assistance with electron microscopy imaging we acknowledge the Scientific Center for Optical and Electron Microscopy (ScopeM, ETH Zurich), as well as the Binnig and Rohrer Nanotechnology Center (BRNC) at IBM Zurich for help with substrate preparation.

# Supporting Information

# Electrochemical 3D Printing of Silver and Nickel Microstructures with FluidFM

Cathelijn van Nisselroy, Chunjian Shen, Tomaso Zambelli, Dmitry Momotenko*

* Correspondence to e-mail; dmitry.momotenko@uol.de

## S1 Print Rate of Ag and Ni for Additional Pressure Ranges

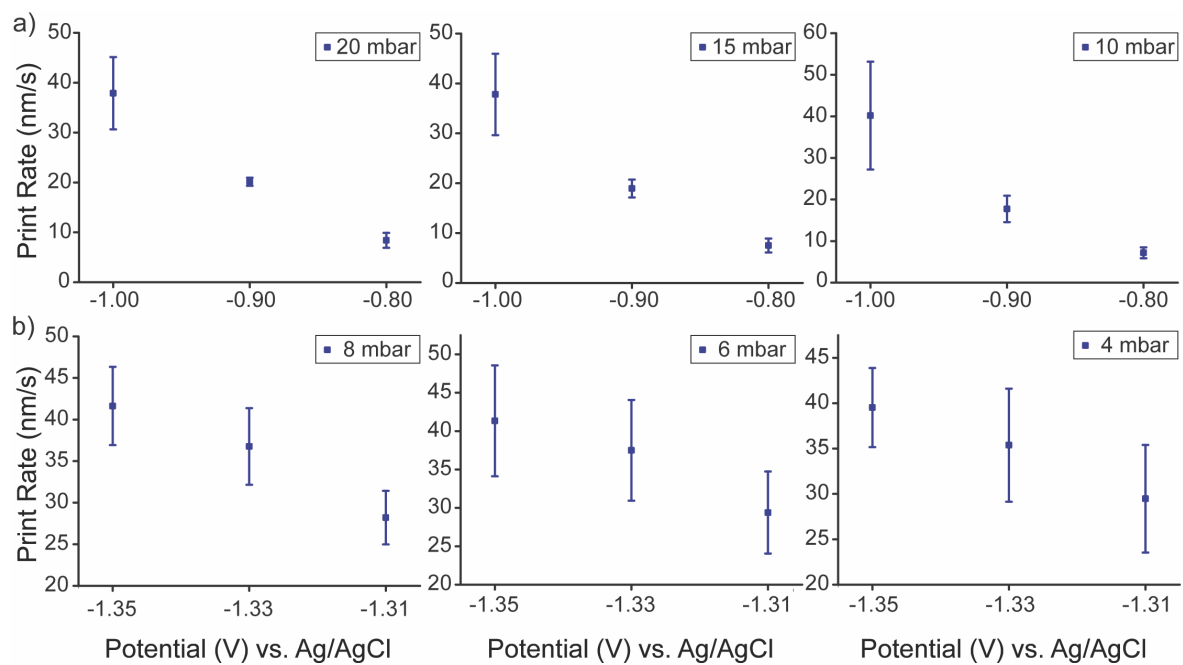

***Figure S1*** Print rate (nm/s) as a function of overpotential, here displayed for the other pressure studied for **a)** Ag and **b)** Ni. The standard deviation and mean values are calculated for $N = 3$.

## S2 Volumetric Throughput of Ag and Ni Pillars for Other Printing Potentials

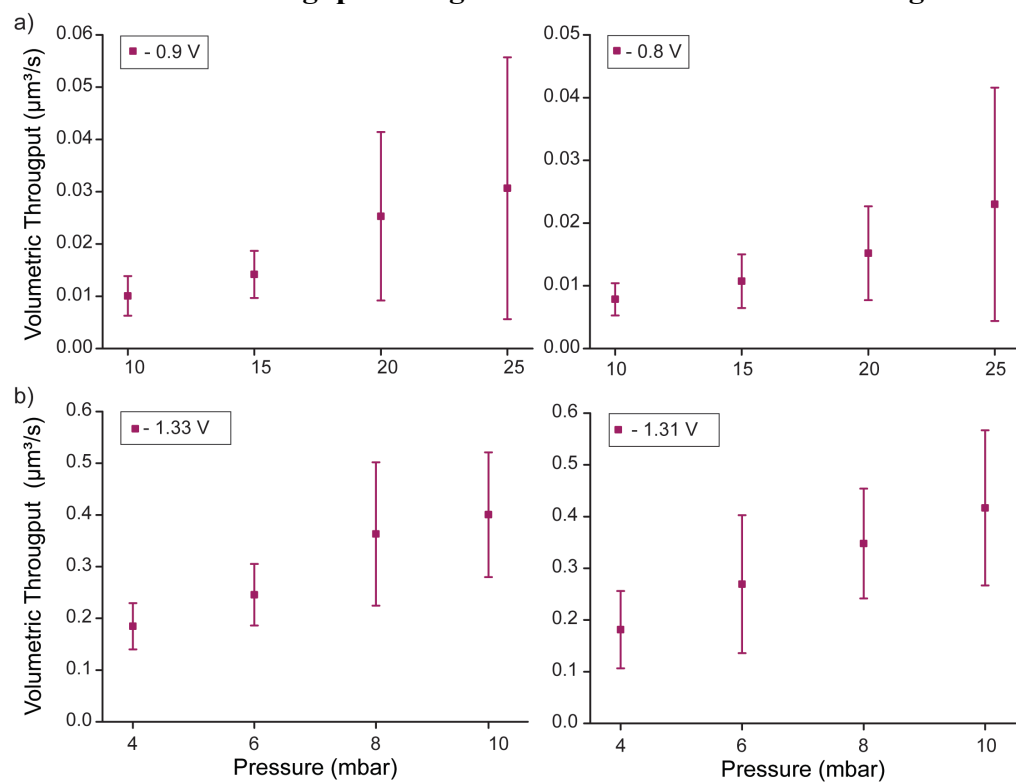

***Figure S2*** Volumetric througput (um$^3$/s) as a function of pressure displayed for the deposition window of **a)** Ag and **b)** Ni. The standard deviation and mean values are calculated for *N* = 3.

## S3 EDX Analysis of Ag and Ni Samples

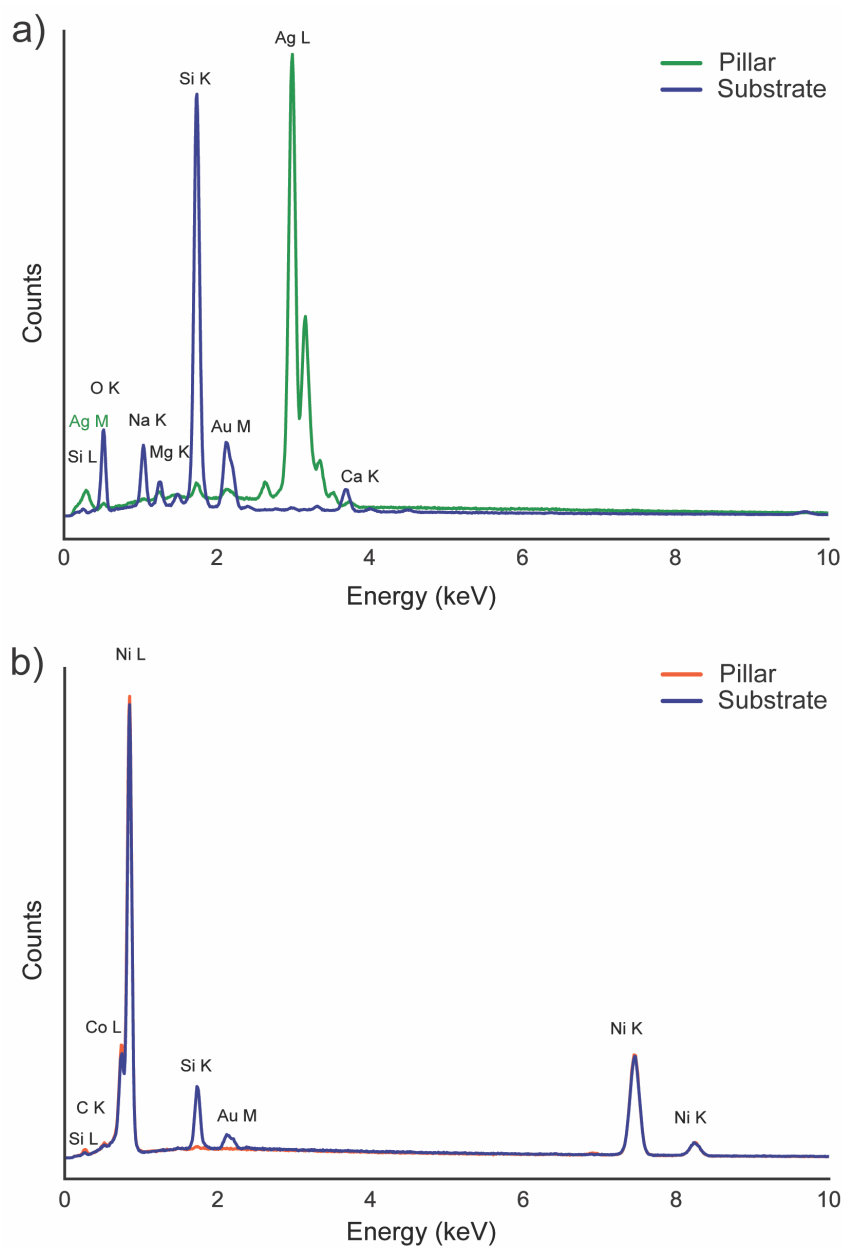

*Figure S3.* EDX spectrum of **a)** an Ag pillar and the substrate, **b)** a Ni pillar and the substrate, to demonstrate the chemical composition. On both spectra only main peaks are labeled.

To confirm the chemical composition of the printed structures, EDX spectra were acquired from both the printed structure and the substrate. Figure S3a displays the spectrum for the Ag printed sample, where the green line represents the spectrum of the pillar and the dark blue line that of the printed pillar. The spectral peaks for Si-L, Si-K and O-K emerge from the glass substrate, whereas the Au-M peak is dedicated to the substrate's coating. Na-K, Mg-K and Ca-K peaks found on the substrate could be attributed to leftover contaminations after the sample washing steps. Ag-M and the three Ag-L peaks are clearly visible in the pillar's spectrum, thereby confirming the Ag nature of the printed pillar. Si-K and Au-M also appear up to a minimum in the sample's spectrum, where most likely a limited part of the substrate was scanned additionally as the sample scan area, the top of the pillar, is rather small.

Figure S3b shows the spectrum for the Ni printed sample. Here, the orange line represents the spectrum of the pillar and the dark blue line that of the substrate. The characteristic Ni spectral peaks can be found in the spectra of both substrate and printed pillar as before printing a thin layer of Ni was electrodeposited on the substrate to facilitate the 3D printing of Ni. Moreover, a clear cobalt peak is detected, which may be ascribed to known Co contaminations in the Ni powder. A carbon peak is present for both the printed pillar and substrate and can be ascribed to carbon contamination, often occurring in electron microscopy and EDX experiments from interaction of electron beam with organic contaminants.